\documentclass[showpacs,prl,a4paper,floatfix,twocolumn]{revtex4}
\usepackage{graphicx}
\usepackage{bm,epsfig}
\usepackage{amsmath}
\usepackage{amssymb}

\begin{document}

\title{Four-wave mixing enhanced white-light cavity}

\author{Robert \surname{Fleischhaker}}
\email{robert.fleischhaker@mpi-hd.mpg.de}
\affiliation{Max-Planck-Institut f\"ur Kernphysik, Saupfercheckweg 1, 
D-69117 Heidelberg, Germany}

\author{J\"org \surname{Evers}}
\email{joerg.evers@mpi-hd.mpg.de}
\affiliation{Max-Planck-Institut f\"ur Kernphysik, Saupfercheckweg 1, 
D-69117 Heidelberg, Germany}

\date{\today}

\begin{abstract}
We discuss in-medium propagation dynamics in a white light cavity that leads to an enhancement of the cavity's bandwidth without reducing its maximum intensity buildup.
We analyze the spatiotemporal dynamics of our system with a full simulation of the field propagation in a regime that leads to strong absorption of the control fields. We find that an additional coherent field is generated within the medium via four-wave mixing. This self-generated field leads to a backaction of the medium onto the probe field. Counter intuitively, this pronounced in-medium dynamics throughout the propagation leads to an additional enhancement of the cavity bandwidth.
\end{abstract}


%
%
%
%
%
%
%

\maketitle

In an optical cavity, the bandwidth of supported frequencies and the intensity buildup are inversely proportional~\cite{saleh}. Increasing the cavity's finesse, e.g., via the reflectivity of the mirrors, leads to a higher buildup for a smaller range of frequencies and vice versa. The reason is that frequencies away from the cavity resonance correspond to different wavelengths which do not exactly fulfill the resonance condition. Thus, they acquire a phase shift with respect to the resonance frequency and experience loss at the mirrors.
In terms of applications, this inverse dependence is a limiting factor for a number of schemes. Perhaps most prominently, gravitational wave detectors (GWD) aim at detecting tiny oscillations that ideally could be amplified by the power buildup in a high-quality cavity with large bandwidth~\cite{revGWD}. To overcome this problem, the concept of a so-called white-light cavity (WLC) was developed~\cite{first_wlc}. Its basic idea is to employ a mechanism inside the cavity that cancels the phase shift for off-resonant frequencies, thereby improving the bandwidth of a cavity without the drawback of reducing its maximum buildup. In the case of GWD one could increase sensitivity without restricting detection bandwidth.

It may seem that the simplest implementation of a WLC would be to introduce a pair of plain parallel gratings such that their diffraction leads to a frequency dependent path length inside the cavity~\cite{grating1}, which, however, is not feasible~\cite{grating2}. A grating causes a phase shift that depends on the position where the wave is diffracted. Since different frequencies are diffracted at different positions at one of the gratings, an additional frequency dependent phase shift occurs and prohibits canceling the frequency dependent phase shift in the cavity.

A different implementation of WLC uses a medium with negative dispersion inside the cavity. In such a medium, phase shifts due to wavelength mismatch can be compensated by suitable phase shifts generated via a frequency-dependent index of refraction.
Proposed systems include a strongly driven double-$\Lambda$ system with incoherent pumping \cite{first_wlc}, a strongly driven two-level atomic resonance, and a $\Lambda$-system off-resonantly driven by two strong fields \cite{twolevel}. In the latter case, the negative dispersion occurs between two gain lines. This has also been used in an experiment to demonstrate negative group velocity \cite{two_gainlines}, a closely connected phenomenon, and recently the first experimental demonstration of a WLC was accomplished in such a system \cite{gainlines_wlc}. In a different experiment the nonlinear negative dispersion occurring in a standard $\Lambda$-system at higher probe field intensities was used \cite{nonlinear_dispersion}. In this case, however, the cavity bandwidth becomes dependent on the probe field intensity.
Complementary to the original WLC approach, recently a high-quality white-light cavity was  demonstrated with a whispering gallery mode resonator~\cite{WGM}, which relies on an effectively continuous mode spectrum, however, with rather low input and output coupling.

The experimental results show that the concept of a cavity bandwidth enhancement with a negative dispersion medium is promising. But whether a real benefit in applications beyond proof of principle will be possible with this concept will depend on the flexibility of the level scheme, scalability of its parameters, and the influence of competing or disturbing processes on the performance. Advancing to more complex level schemes than standard electromagnetically induced transparency (EIT) based setups motivated by the desire for better control over the WLC, however, typically leads to absorption in the probe or control field amplitudes throughout the propagation, severely degrading the performance in the required extended media.

\begin{figure}[t]
\includegraphics[width=8cm]{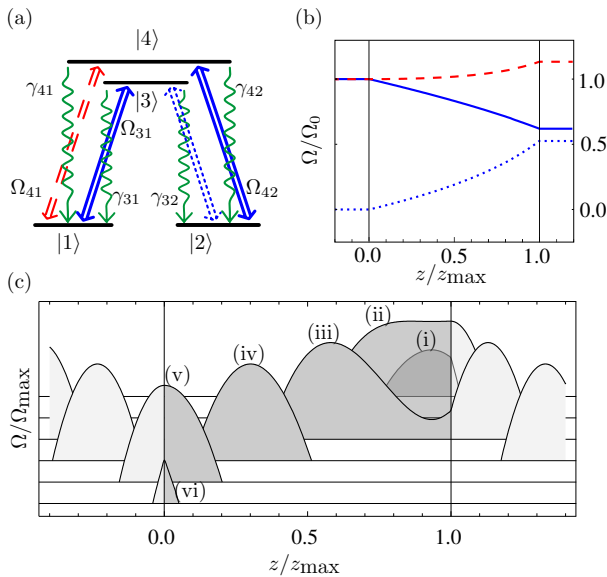}
\caption{\label{fig1}(Color online)
(a) The considered double-$\Lambda$ level scheme. Two transitions are driven by strong continuous-wave fields indicated by the thick solid blue arrows. One transition couples to the probe field indicated by the dashed red arrow. The dotted blue arrow represents a field generated within the medium via four-wave mixing. $\Omega_{jk}$ are Rabi frequencies. The spontaneous decays with rates $\gamma_{jk}$ are denoted by the wiggly green lines ($j\in \{3,4\}$, $k\in \{1,2\}$). (b) Attenuation or amplification of the control fields (solid blue line), probe field (dashed red line), and generated field (dotted blue line) over the medium. The generated field is scaled to the initial value of the probe field. (c) Peak of a probe field pulse for successive times (i)-(vi), scaled to the respective maximum in time. The peak leaves the medium before it enters and runs through the medium in reversed direction.
}
\end{figure}

In this Letter, we discuss such propagation effects in strongly absorbing WLC media, and show that instead, the WLC performance can be improved due to an electromagnetic field that is generated and sustained by the atomic medium itself via four-wave mixing (4WM). 
For this, we study light propagation through an atomic four-level medium in double-$\Lambda$ configuration as depicted in Fig.~\ref{fig1}(a). The medium is prepared by two control fields coupling to the transitions $|1\rangle\leftrightarrow |3\rangle$ and $|2\rangle\leftrightarrow |4\rangle$.
If in addition the probe field is applied to the $|1\rangle\leftrightarrow |4\rangle$ transition, then 4WM becomes possible, and a new field is generated within the medium on the transition $|2\rangle\leftrightarrow |3\rangle$. With this additional 4WM field present, the applied fields form a closed interaction loop that allows for scattering processes back into the probe field mode. Thus, a backaction of the field generated via 4WM out of the probe and coupling fields on the probe field itself occurs, controlled by the additional coupling fields. It is obvious that these processes occur during the propagation of the light through the medium and cannot be captured  in a standard treatment in terms of a single atom susceptibility analysis. Also, the two coupling fields are absorbed such that their intensity changes substantially throughout the medium, at the same time preparing the medium in a position-dependent initial state. A typical modification of the field amplitudes in the medium is shown in Fig.~\ref{fig1}(b). Therefore, apart from theoretical modelling, we numerically study the full propagation dynamics of all fields through the medium in order to determine their influence on the performance of a WLC.

Before we present our actual results we start with discussing an empty cavity resonance profile which is given by 
 $I = I_{max}/[1 + (2 \mathcal{F} / \pi)^2 \textrm{sin}^2(\varphi / 2)]$,
where $I_{max} = I_0 / (1 - r)^2$ is the maximum intensity buildup, $\mathcal{F} = \pi r^{1 / 2} / (1 - r)$ is the cavity finesse, and $r e^{i \varphi}$ is the round trip loss and phase shift~\cite{saleh}. Without a medium the phase shift with respect to the resonance frequency is given by
$\varphi_0 = 2 L \Delta/c$,
with $L$ the cavity length, $\Delta = \omega - \omega_0$ the detuning from the resonance frequency $\omega_0$, and $c$ the speed of light in vacuum. This phase shift leads to a cavity bandwidth (FWHM) of
\begin{align}
 \label{fwhm0}
 \gamma_0 = \frac{2 \pi c }{2 L\mathcal{F}}.
\end{align}
A medium of length $l < L$ and refractive index $n$ inside the cavity leads to an additional phase shift
$\varphi_1 = 2 l \omega_0 (n - 1)/c$.
We assume that close to the resonance frequency $n$ can be approximated as 
\begin{align}
 \label{refr}
 n = 1 + \frac{n_g}{\omega_0} \Delta + n_3 \Delta^3 + \mathcal{O}(\Delta^4),
\end{align}
where $n_g = \omega_0 \frac{\partial n}{\partial \omega} \big\vert_{\omega_0}$ is the group index and $n_3 = \frac{1}{6} \frac{\partial^3 n}{\partial \omega^3} \big\vert_{\omega_0}$ is the third order correction term to a linear slope. A second order term does not appear since the resonance frequency is an inflection point of the dispersion if probe field frequency and cavity resonance  coincide. From the condition $\varphi_0 + \varphi_1 = 0$ for a WLC we find
\begin{align}
 \label{wlcc}
  n_g = - \frac{L}{l}.
\end{align}
Now the terms linear in frequency cancel and the enhanced cavity bandwidth is given by (FWHM)
\begin{align}
 \label{fwhm1}
 \gamma_1 = \left( \frac{4 \pi c }{ l \omega_0 n_3\mathcal{F}} \right)^\frac{1}{3}.
\end{align}

From the WLC condition (Eq.~\ref{wlcc}) we see that a WLC requires a medium with negative group velocity. Therefore, in a first calculation, we propagate probe pulses of different bandwidths through the double-$\Lambda$ medium and optimize parameters for a negative group index. For this, we derive the Maxwell-Schr\"odinger equations describing the light propagation through the medium using standard techniques~\cite{propagation}, and solve the equations numerically on a grid using a Lax-Wendroff integration method~\cite{wendroff}. 
We assume a $l = 0.3$\,m long medium with a density of $N = 6.6 \times 10^{15}\,\textrm{m}^{-3}$ atoms and initial control field strengths of $\Omega_{42} = 15.5\,\gamma$ and $\Omega_{31} = 16\,\gamma$. 
The weak probe field ($\Omega_{41} = 0.1\,\gamma$) is applied to transition $|1\rangle \leftrightarrow |4\rangle$.
For a Gaussian probe field envelope the peak of the pulse leaves the medium before it enters and runs through the medium in reversed direction (see Fig.~\ref{fig1} (c)). This behavior is typical for a medium with negative group velocity \cite{two_gainlines, speedreview}. From the advancement in time $T_a$ of the pulse peak after passing the medium we calculate the group index,
 $n_g = - c \, T_a/l$.

As we are most interested in propagation effects, the evolution of the different field amplitudes throughout the medium is shown in Fig.~\ref{fig1} (b).
The control fields are attenuated to about $60\%$ of their initial value. At the same time, the control fields together with the probe field generate an additional field on the transition $|2\rangle\leftrightarrow |3\rangle$ via 4WM. 
This internally generated field and the externally applied fields form a closed interaction loop~\cite{closed_loop_0,closed_loop, closed_loop_2}, that in turn leads to light scattering into the probe field mode. By means of this backaction, the probe field is amplified by about $10\%$, which allows to compensate for losses inevitable in an experimental realization.

\begin{figure}[t]
\includegraphics[width=8cm]{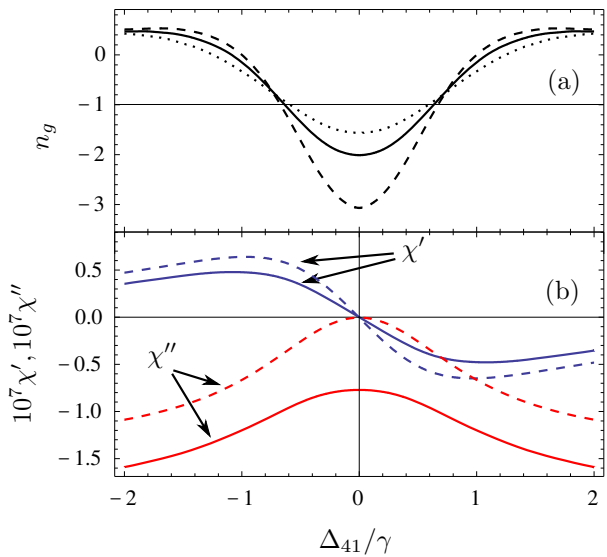}
\caption{\label{fig2}(Color online)
(a) Group index of a Gaussian probe pulse with bandwidth of $2 \gamma$ (dotted line) and $\gamma / 2$ (solid line). The dashed line shows bandwidth $\gamma / 2$, but with field generation via 4WM suppressed. Bandwidth narrowing below $\gamma / 2$ does not change the group index significantly.  (b) Real part $\chi'$ (upper blue lines) and imaginary part $\chi''$ (lower red lines) of the effective probe field susceptibility with (solid lines) and without (dashed lines) 4WM.
}
\end{figure}

In order to evaluate the medium performance, in a second calculation, we extract the effective susceptibility of the medium by comparing the amplitude and the phase of a continuous wave probe field at the medium entry and exit after full numerical propagation, simulating an experimental measurement.
For this, we relate the probe field at the medium exit $\Omega_{41}(l)$ to the field at the entry $\Omega_{41}(0)$ by
\begin{align}
 \Omega_{41}(l) = \Omega_{41}(0) e^{- k l \frac{\chi''}{2}} e^{i k l \frac{\chi'}{2}},
\end{align}
where $k$ is the respective wave number whereas $\chi'$ and $\chi''$ are the real and imaginary part of the susceptibility.
The obtained susceptibility is a measure for the effect the medium would have on the probe field during each passage through the cavity. 

\begin{figure}[t]
\includegraphics[width=8cm]{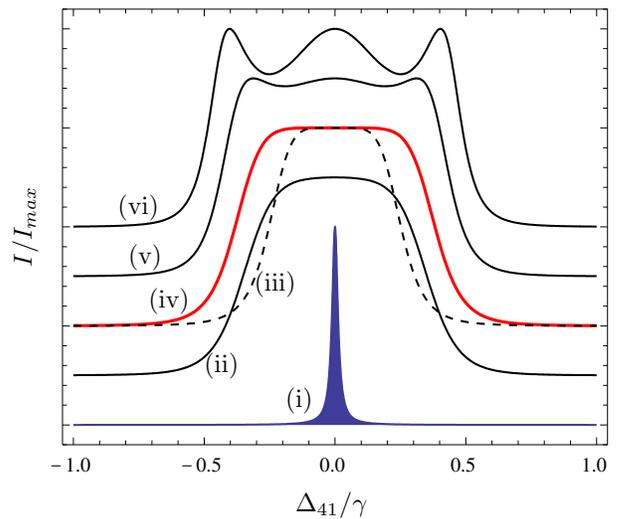}
\caption{\label{fig3}(Color online)
Intensity buildup for an $\mathcal{F} = 1000$ cavity without medium (i) and with the proposed WLC medium (iv). Curve (iii) shows the corresponding medium result with four-wave mixing artificially suppressed. The other three profiles show the general case with 4WM and a $2 \%$ (ii), a $-2 \%$ (v), and a $-4 \%$ (vi) mismatch of the cavity length with respect to the WLC condition. For better visibility the different profiles have been shifted by multiples of $0.25$ with respect to each other. Profiles (iii) and (iv) have the same shift.
}
\end{figure}

The results for the group index and the effective susceptibility are shown in Fig.~\ref{fig2} against the probe field detuning. In addition, we also show as dashed lines the corresponding susceptibility that is obtained if the medium backaction via 4WM and the closed-loop scattering is suppressed. This is achieved by artificially setting the generated field on transition $|2\rangle\leftrightarrow |3\rangle$ to zero throughout the numerical analysis.
In the latter case, the effective susceptibility can perfectly be explained by averaging the result of a single-atom analysis over the non-uniform control field intensities in the medium. 
In contrast, in the general case with 4WM and medium backaction present, a single-atom analysis combined with an averaging fails to give the correct results.

It can be seen from Fig.~\ref{fig2} that the probe field amplification via the 4WM-generated control field leads to a less steep slope and a smaller group index (Fig.~\ref{fig2} (a) and (b), solid lines). It may seem that a smaller group index deteriorates the bandwidth enhancement. But we see that as long as the group index for a resonant pulse reaches values below $n_g = -1$, the WLC condition Eq.~(\ref{wlcc}) can still be fulfilled. Only the ratio of medium length to cavity length has to be adjusted. In our numerical calculation, we find $n_g \simeq -2$ which corresponds to a medium filling about half of the cavity.

In Fig.~\ref{fig3}, we finally show the enhanced cavity resonance profile. We assume a cavity finesse of $\mathcal{F} = 1000$ which corresponds to a mirror reflectivity of $r = 99.68\%$, and a cavity length of $L=59.5$\,cm for which the WLC condition is fulfilled in the case with 4WM. With 4WM suppressed, the negative group index is larger such that the WLC condition is violated for the same parameters. We compensate this by adjusting the medium length $l$ suitably, but keep the cavity length $L$ fixed such that the empty cavity bandwidth $\gamma_0$ is equal in both cases.
Without 4WM, we find a cavity bandwidth enhancement by about a factor of $20$ (Fig.~\ref{fig3} (iii)). With 4WM, the enhancement is by a factor of about $30$ (Fig.~\ref{fig3} (iv)). Thus, the in-medium dynamics enhances the desired bandwidth increase.

At the WLC, the enhanced bandwidth profiles become quadratically flat around the resonance frequency. Under- or overcompensating the frequency dependent phase shift leads to a less flat response, see Fig.~\ref{fig3}.

\begin{figure}[t]
\includegraphics[width=8cm]{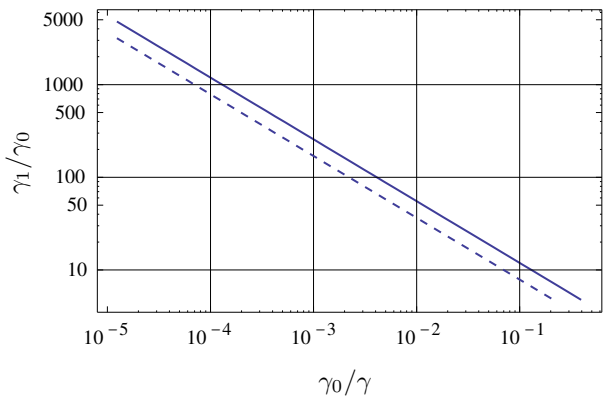}
\caption{\label{fig4}(Color online)
Enhancement factor for the considered WLC scheme (solid line) and with additional 4WM-field suppressed (dashed line) against empty cavity bandwidth $\gamma_0$.}
\end{figure}

We also calculated the cavity bandwidth enhancement for different values of empty cavity bandwidth and for both cases, with and without 4WM. The result is shown in Fig.~\ref{fig4} in a double-logarithmic plot. Both curves are indistinguishable from the theoretical power law
\begin{align}
 \gamma_1/\gamma_0 \propto \left( \gamma_0/\gamma \right)^{-\frac{2}{3}}
\end{align}
that is evident from Eqs.~(\ref{fwhm0}) and (\ref{fwhm1}) which shows that higher order terms in Eq.~(\ref{refr}) do not play a significant role. Again, in Fig.~\ref{fig4}, the bandwidth with full medium dynamics (solid line) is enhanced compared to the case without 4WM (dashed line).

From Fig.~\ref{fig4}, bandwidth enhancement factors above $10^3$ are theoretically predicted for high quality cavities. This result, however, neglects practical issues common to light propagation setups with gases as follows.
In an atomic gas medium suitable for the required phase compensation, typically Doppler effects need to be considered. In the existing experiments, this issue could be overcome. As far as the propagation effects in our level scheme are concerned, the relevant processes are two-photon Raman transitions and four-photon closed-loop transitions. For these transitions the Doppler effect typically cancels to first order at the resonance frequency
if the fields co-propagate. 
However, since the mechanism for a WLC relies on a frequency range around the resonance frequency, quantitative changes to our results can be expected. 
Second, an exact fulfilling of the WLC condition requires a stabilization of the different parameters such as the control field strengths and the cavity length. An accurate control of the cavity length e.g. via piezo elements could also be used to compensate fluctuations in the control field strengths. From Fig.~\ref{fig3}(ii), (v), and (vi) it can be seen that a WLC condition mismatch on the few percent level typically is not critical.
As far as the application of a WLC in a GWD is concerned in principle the bandwidth enhancement factor should be directly convertible into a sensitivity enhancement of the same order. A question that has to be addressed, however, is how to actually implement a WLC into a GWD. Especially the larger scales of the cavity in the case of existing GWDs \cite{revGWD} consisting of the interferometer and a so-called signal recycling mirror pose a demanding task.

In summary, we have investigated a white light cavity enhanced by in-medium propagation dynamics. An additional light field is generated via four-wave mixing during the light propagation, which leads to a closed-loop configuration of the applied coherent fields. This closed loop configuration allows for a backaction on the probe field mode, giving rise to small gain that could be used to compensate other loss channels, and at the same time enhances the bandwidth increase. More generally, our results show that setups with strong 
in-medium dynamics could be an interesting alternative
to the existing EIT-based light propagation schemes.

\begin{acknowledgments}
JE would like to thank M. S. Zubairy for helpful discussion.
\end{acknowledgments}



\begin{thebibliography}{99}

\bibitem{saleh}
B. E. A. Saleh and M. C. Teich, 
{\em Fundamentals of Photonics} 
(Wiley, New York, 1991).

\bibitem{revGWD}
P. Aufmuth and K. Danzmann,
New J. Phys. {\bf 7}, 202 (2005).

\bibitem{first_wlc}
A. Wicht  {\it et al.}, 
Opt. Commun.  {\bf 134}, 431 (1997).

\bibitem{grating1}
S. Wise  {\it et al.}, 
Classical Quantum Gravity {\bf 21}, S1031 (2004).

\bibitem{grating2}
S. Wise {\it et al.}, 
Phys. Rev. Lett. {\bf 95}, 013901 (2005).


\bibitem{twolevel}
R. H. Rinkleff and A. Wicht, 
Phys. Scr. T {\bf 118}, 85 (2005).

\bibitem{two_gainlines} 
A. Dogariu, A. Kuzmich, and L. J. Wang, 
Phys. Rev. A {\bf 63}, 053806 (2001).

\bibitem{gainlines_wlc}
G. S. Pati, M. Salit, K. Salit, and M. S. Shahriar, 
Phys. Rev. Lett. {\bf 99}, 133601 (2007).

\bibitem{nonlinear_dispersion}
H. Wu and M. Xiao, 
Phys. Rev. A {\bf 77}, 031801 (R) (2008).

\bibitem{WGM}
A. A. Savchenkov, A. B. Matsko, and L. Maleki, 
Opt. Lett. {\bf 31}, 92 (2006).

\bibitem{closed_loop_0}
S. J. Buckle  {\it et al.}, 
Opt. Acta {\bf 33}, 1129 (1986).

\bibitem{closed_loop}
M. Mahmoudi and J. Evers,
Phys. Rev. A {\bf 74}, 063827 (2006).

\bibitem{closed_loop_2}
R. Fleischhaker and J. Evers,
Phys. Rev. A {\bf 77}, 043805 (2008).

\bibitem{propagation}
M. O. Scully and M. S. Zubairy, {\it Quantum Optics} (Cambridge University Press,
Cambridge, 1997).

\bibitem{wendroff}
W. H. Press {\it et al.}, 
{\it Numerical Recipes} 
(Cambridge University Press, Cambridge, 1992).

\bibitem{speedreview}
P. W. Milonni, 
{\em Fast Light, Slow Light and Left-Handed
Light} (Institute of Physics, 2005).



\end{thebibliography}
\end{document}